\documentclass[]{spie}  

\usepackage{amsmath,amsfonts,amssymb}
\usepackage{rotating}
\usepackage{graphicx}
\usepackage[colorlinks=true, allcolors=blue]{hyperref}

\title{SuMAC: On-sky demonstration of multi-pixel on-chip millimeter-wave spectroscopy at the Large Millimeter Telescope}

\author[a]{Marcial Becerril-Tapia}
\author[a]{Peter S. Barry}
\author[a]{Christopher S. Benson}
\author[b]{Charles M. Bradford}
\author[a]{Thomas L. R. Brien}
\author[c]{Scott C. Chapman}
\author[a]{Simon Doyle}
\author[d]{Víctor Gómez-Rivera}
\author[e]{Steven Hailey-Dunsheath}
\author[d]{Jesús Hernández-Aguilar}
\author[d]{José Luis Hernández-Rebollar}
\author[d]{David H. Hughes}
\author[e]{Elijah Kane}
\author[f]{Kirit S. Karkare}
\author[f]{Alex M. Lapuente}
\author[g]{Ryan McGeehan}
\author[d]{Raúl Naranjo-Romero}
\author[a]{Andreas Papageorgiou}
\author[h]{Joseph Redford}
\author[d]{Iván Rodríguez-Montoya}
\author[a]{Sam Rowe}
\author[d]{David O. Sánchez-Argüelles}
\author[f]{Sofia Savorgnano}
\author[f]{Annie Tan}
\author[d]{Jair Vega-Méndez}

\affil[a]{Cardiff University, 5 The Parade, Cardiff, CF24 3AA, United Kingdom}
\affil[b]{Jet Propulsion Laboratory, 4800 Oak Grove Drive, La Cañada Flintridge, CA 91011, United States}
\affil[c]{Dept. of Physics and Atmospheric Science, Dalhousie University, NS, Canada}
\affil[d]{Instituto Nacional de Astrofísica, Óptica y Electrónica, Luis Enrique Erro 1, Santa María Tonantzintla, Puebla, CP 72480, Mexico}
\affil[e]{California Institute of Technology, 1200 E California Blvd, Pasadena, CA 91125, United States}
\affil[f]{Boston University, 590 Commonwealth Ave, Boston, MA 02215, United States}
\affil[g]{The University of Chicago, 5801 S Ellis Ave, Chicago, IL 60637, United States}
\affil[h]{University of California Santa Barbara, 552 University Road, Santa Barbara, CA 93106, United States}

\authorinfo{Further author information: (Send correspondence to Marcial Becerril-Tapia)\\Marcial Becerril-Tapia: E-mail: Becerril-TapiaM@cardiff.ac.uk, Telephone: +44 7562902823}

\begin{document} 
\maketitle

\begin{abstract}
The SuperSpec-MUSCAT Collaboration (SuMAC) is an effort to operate  SuperSpec spectrometers in the MUSCAT instrument at the Large Millimeter Telescope (LMT). SuperSpec is an on-chip filterbank spectrometer using kinetic inductance detectors (KIDs), with moderate $R\sim 200$ spectral resolution for observing in the 1 mm atmospheric window (190-300 GHz). While nominally a continuum camera operating a 1.1 mm, MUSCAT was also 
designed to operate as a testbed for pioneering detection technologies. In July 2025, three SuperSpec devices were installed and commissioned in the MUSCAT cryostat with minimal modifications to the system. This work presents an overview of SuperSpec's deployment and performance in MUSCAT along with the main results from the Summer 2025 observation campaign. Highlights include the detection of CO emission line in the galaxy NGC253, mapping the Orion KL region, and preliminary line intensity mapping scans.
\end{abstract}


\section{INTRODUCTION}
\label{sec:intro}  

It is well established that approximately half of all radiation in the Universe produced by galaxies at all redshifts lies in the infrared-millimetre range. Most of this energy originates from the absorption of optical/ultraviolet radiation by cosmic dust, which then re-emits it in the infrared\cite{Lagache_2005}. In addition to the characteristic spectral signature of dust continuum emission, this radiation also excites far-infrared line emission, such as the CO($J\rightarrow J-1$) rotational ladder and the [CII] ionised carbon fine-structure transition. Detecting these spectral lines is essential for estimating galaxy redshifts, studying global star formation rates and molecular gas content over cosmic time, and developing line intensity mapping (LIM) of CO and [CII], which involves detecting integrated spectral line emission from galaxies and the intergalactic medium to study the growth and evolution of cosmic structure\cite{kovetz2017lineintensitymapping2017status}, a process crucial for investigating large-scale structures, the Epoch of Reionisation, and the validation of cosmological models.

Although instruments like ALMA have detected these lines, their role is limited to a modest bandwidth and a relatively narrow field of view (FOV). To complement their observations, large-scale instruments with moderate mm-wave spectral resolution and hundreds of detectors are required. Of particular note is the development of on-chip spectrometers based on kinetic inductance detector (KID) technology, in which the dispersion and detection elements are integrated into a single wafer, offering a highly scalable option. This technology has been validated on-sky by pioneering instruments such as DESHIMA\cite{Endo_2019} and SPT-SLIM\cite{Karkare_2022, dibert2025}. SuperSpec—an on-chip, ultra-compact, thin-film lithographic spectrometer—has been developed over the past decade as part of this effort. 

In Summer 2025, an upgrade to the MUSCAT millimetre-wave continuum camera installed on the Large Millimeter Telescope (LMT) provided a window of opportunity to install and operate the SuperSpec detectors in its focal plane. With relatively minimal modifications, we were able to commission the instrument and carry out a brief observation campaign during Summer 2025, focusing on instrument characterisation, on-sky validation of the on-chip spectrometer technology, and the assessment of opportunities at the telescope for future improvements and developments. 

In this work, we present the preliminary results from SuperSpec on MUSCAT, known as SuMAC (SuperSpec-MUSCAT Collaboration), during that campaign. We describe the operation and performance, briefly outlining the key modifications made to adapt them to one another. We present a summary of the observations conducted for calibration, characterisation, and spectral verification. Based on these observations, we present the on-sky instrument performance, including beam shapes and sensitivities expressed as Noise Equivalent Temperature (NET). We highlight spectral observations of the galaxy NGC 253 with a detection of the CO($J=2\rightarrow1$) line at 230.5 GHz, as well as a multispectral map of the central region of the Orion Molecular Cloud. Finally, we present a preliminary evaluation of the results and future work.

Four other proceedings accompany this work, presenting more details on the SuMAC deployment. Characterisation of the SuperSpec filter banks is presented in Kane et al. (2026)\cite{Kane2026}, flux calibration and data reduction in Lapuente et al. (2026)\cite{Lapuente26}, noise and common mode analysis in Redford et al. (2026)\cite{Redford2026}, and Savorgnano et al. (2026)\cite{Savorgnano26} present the preliminary analysis of the first LIM-like observations carried out with SuMAC.

\section{SuMAC: The SuperSpec-MUSCAT collaboration}

The SuMAC focal plane consisted of three SuperSpec on-chip filter-bank spectrometers. The instrument was sensitive to 190--300 GHz with a spectral resolution of R$\sim200$.  
Each pixel contains a silicon hemispherical lens that couples radiation to a broadband dual-slot antenna that feeds  a lithographically-patterned filter bank of half-wave microstrip resonators. Each resonator absorbs its corresponding spectral fraction of the radiation, according to its resonant frequency and coupling quality factor (dispersion element). Finally, a KID made up of a TiN inductor and Nb inter-digital capacitor terminates the output of each filter, absorbing the filtered radiation\cite{Redford_2022}. 

The Mexico-UK Submillimetre Camera for Astronomy (MUSCAT) is a 1.1 mm continuum camera featuring 1,458 dual-polarisation, lumped-element kinetic inductance detectors (LEKIDs) grouped into six sub-arrays~\cite{Tapia2024}. The camera is installed and operational at the LMT, a 50-meter single-dish telescope, located at an altitude of 4,600~m on the summit of the Sierra Negra volcano in Mexico. The detectors operate at a base temperature of 100 mK using a miniature dilution refrigerator\cite{Brien2020}. As part of its science goals, MUSCAT will provide high-resolution follow-up surveys of galactic and extragalactic sources previously observed with the Herschel Space Observatory. 

\begin{figure} [ht]
\begin{center}
\begin{tabular}{c}
\includegraphics[height=6.9cm]{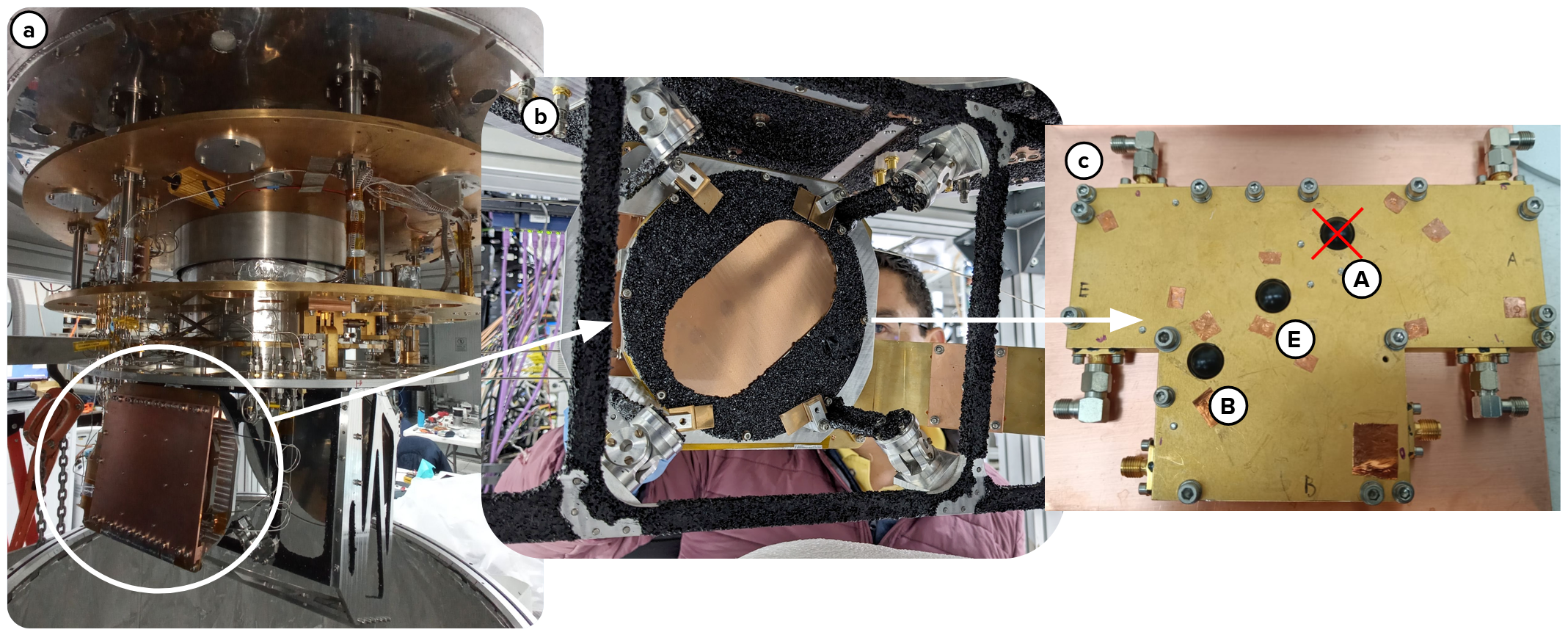}
\end{tabular}
\end{center}
\caption[example] 
{ \label{fig:fig1} a) Interior of the MUSCAT cryostat with the SuperSpec array mounted at its focal plane (white circle). b) Front view of the MUSCAT focal plane, showing the SuperSpec filter stack (high-pass filter and polariser) and—faintly visible through them—the three microlenses of each SuperSpec pixel. c) SuperSpec detector array for SuMAC. Pixels E and B contain 110 channels ranging from 190 to 300 GHz, whereas pixel A has 50 channels. The latter pixel is excluded from most of the measurements presented in these proceedings.}
\end{figure}

Furthermore, MUSCAT is also designed to serve as a testbed for the development of detectors and readout systems. Exploiting this property and its availability during the Summer of 2025, we mounted the SuperSpec array at the focal plane of MUSCAT (see Figure~\ref{fig:fig1}). The SuperSpec array used for the MUSCAT experiment consists of three pixels as shown in Figure~\ref{fig:fig1}c. The central pixel—designated as device E—and one of the side pixels (device B) feature 110 channels (190-300 GHz), whereas the remaining side pixel has 50 channels\cite{Redford_2022}. Preliminary results presented here focus on the central pixel, with the performance of the other pixels, as well as the combined response, will be presented in future publications.

For the assembly, adaptation, and operation of SuperSpec in the MUSCAT instrument (Figure \ref{fig:fig2}), the following relatively minimal modifications were made:

\begin{figure} [ht]
\begin{center}
\begin{tabular}{c}
\includegraphics[height=10cm]{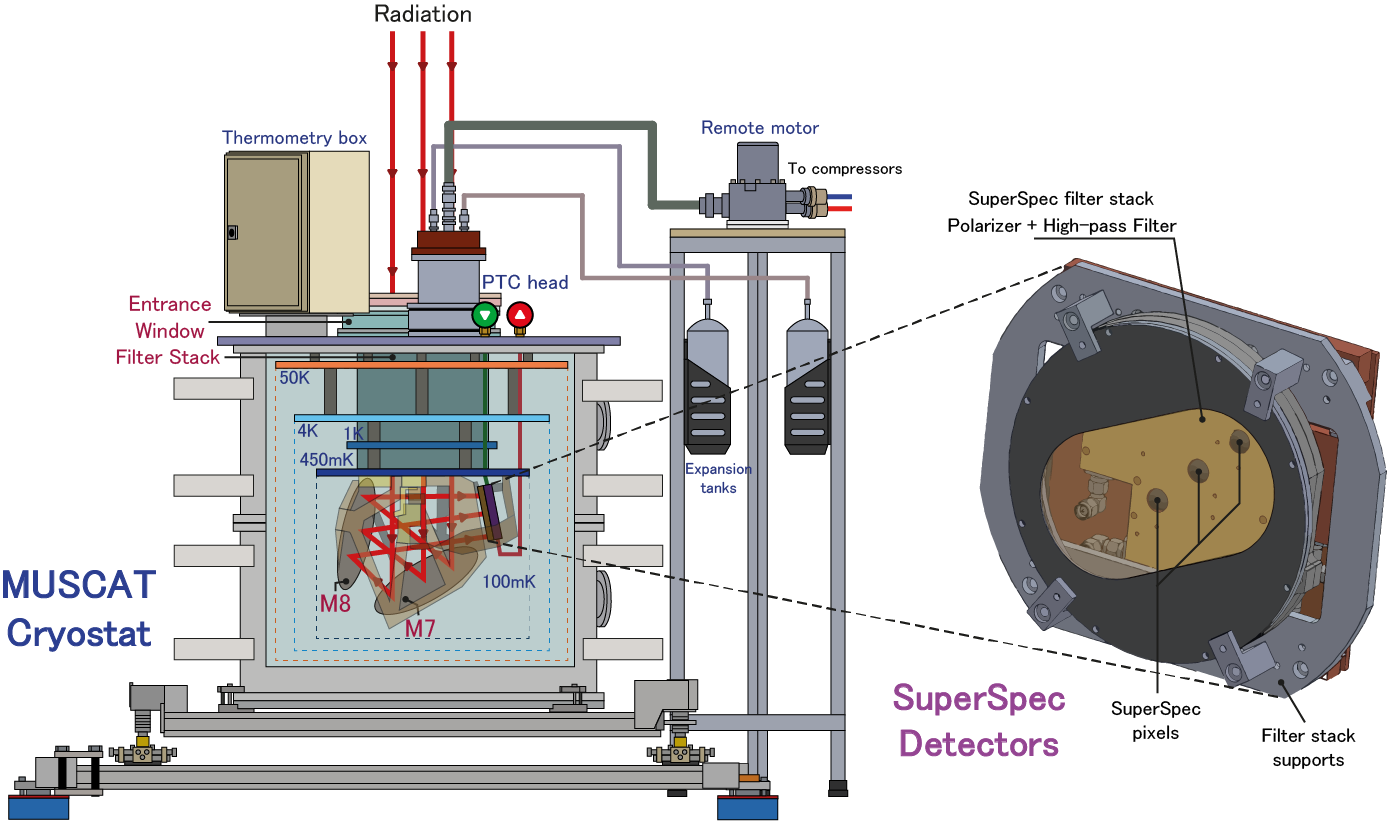}
\end{tabular}
\end{center}
\caption[example] 
{ \label{fig:fig2} SuMAC operational schematic. Radiation collected by the telescope enters the MUSCAT cryostat window and passes through several internal metal-mesh low-pass filters before reaching the innermost cold mirror pair (M7/M8) in a crossed Dragone configuration; these reflect the radiation toward the focal plane, where a metal-mesh high-pass filter defines the instrument's passband, and the radiation finally is absorbed by the three pixels of the SuperSpec array (detail view). The detectors are held at $\sim$95 mK by a miniature dilution refrigerator developed by Chase Research Cryogenics.}
\end{figure} 

\begin{itemize}
    \item The upper bandwidth limit of MUSCAT, 300 GHz, is defined by a metal-mesh low-pass filter installed along MUSCAT's internal optical path before the beam reaches mirrors M7–M8 (see Figure \ref{fig:fig2}), while the lower limit, $\sim$250 GHz, is defined by the waveguide output of the feedhorn block mounted over the detectors\cite{tapia2020muscatfocalplaneverification}. For SuperSpec, we used the existing filter stack and added a metal-mesh high-pass filter to define the low frequency bandpass, ranging from $\sim190$ to 300 GHz. A pair of aluminum aperture plates, each featuring a central oval slot, were fabricated to hold the filter, along with a wire grid polariser, in front of the detector array, as illustrated in the schematic in Figure \ref{fig:fig2} and the photograph in Figure \ref{fig:fig1}b. The outer surface of the aperture stack is blackened using epoxy resin and charcoal to minimise stray reflections (see Figure \ref{fig:fig1}b). An aluminium frame holds the SuperSpec array and mounts it to the MUSCAT focal plane plate.

    We preserve the configuration of warm mirrors that guide and couple the telescope beam \cite{TapiaThesis}.

    \item The readout system in MUSCAT is optimised to operate in the 500 to 1000 MHz range, in contrast to SuperSpec's readout bandwidth of 80 - 450 MHz. While most elements in the MUSCAT readout circuit can extend their operation to lower frequencies, this is not the case for the cryogenic low-noise amplifiers (LNA), whose gain drops sharply below 500 MHz. Consequently, we replaced them with amplifiers with the appropriate bandwidth, developed by Arizona State University (ASU). 

    As part of the planned upgrade to MUSCAT, we removed all DC blocks and replaced the coaxial cables featuring soldered connections with ones using crimped connections. To read out the SuperSpec array, we utilised three of the six readout channels available with MUSCAT—one for each pixel. Figure~\ref{fig:fig3} shows the electrical schematic inside the cryostat for each readout channel. At the input, there is a total attenuation of 30 dB: 20 dB at 4 K, and 10 dB at the intermediate 450 mK stage. At the output, the signal is amplified by approximately 30 dB by the LNAs attached to the 4 K plate. The net gain between input and output is primarily determined by cable loss, approximately -5 dB. 

   \begin{figure} [ht]
   \begin{center}
   \begin{tabular}{c}
   \includegraphics[height=5.5cm]{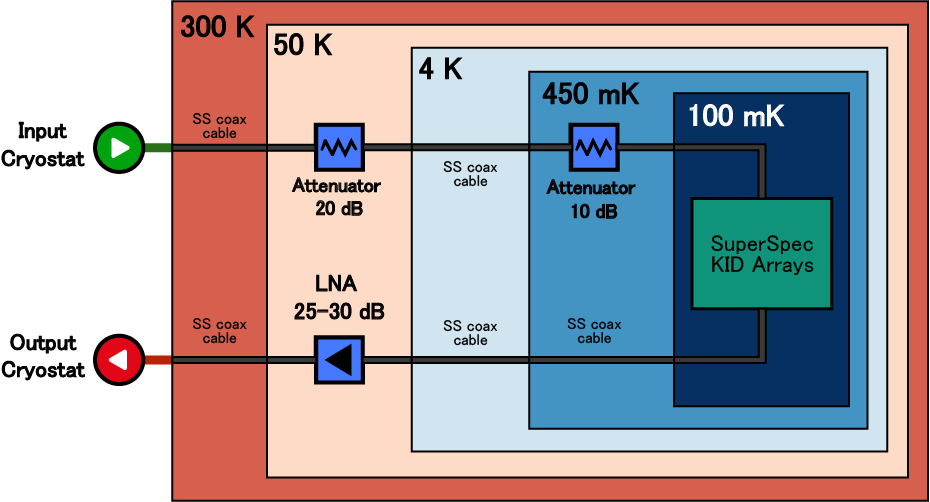}
   \end{tabular}
   \end{center}
   \caption[example] 
   { \label{fig:fig3} Electrical schematic of a cold electronic channel for SuMAC. There is a total of 30 dB of attenuation at the detector array input—20 dB at 4 K and 10 dB in the intermediate stage—to reduce Johnson noise from the warmer stages. At the output, the signal is amplified by LNAs at 4 K.}
   \end{figure} 

    \item For the readout system we combined the existing MUSCAT infrastructure with elements of the native SuperSpec system. Figure \ref{fig:fig4} details the connection diagram for the SuMAC readout system. We retained the MUSCAT Digital Tone Processor\cite{Sam2023}, responsible for generating, reading, and processing the tone comb for each detector via the ROACH-2 board, as well as performing high-frequency analog-to-digital conversions using the MUSIC board. The 512 MHz clock signal is provided by the SuperSpec CLK Generator, utilising the telescope's 10 MHz reference signal. 
    
    For the up- and down-conversion stages, we used the native SuperSpec system that is made up of similar components as the MUSCAT readout adjusted for lower frequency, and employed the local oscillator from MUSCAT.
    A warm amplification stage precedes the cryostat to control the input signal level into the cryostat via a combination of amplifiers and variable attenuator (up to 30 dB).A fixed 20 dB attenuator and a 450 MHz low-pass filter was added to set the required readout power level. At the cryostat output, we included an 800 MHz low-pass filter. Similarly, the output signal is also amplified by a combination of two 40 dB gain amplifiers with an intermediate variable attenuator to regulate the input power to the readout system. 
\begin{figure} [ht]
\begin{center}
\begin{tabular}{c}
\includegraphics[height=6.05cm]{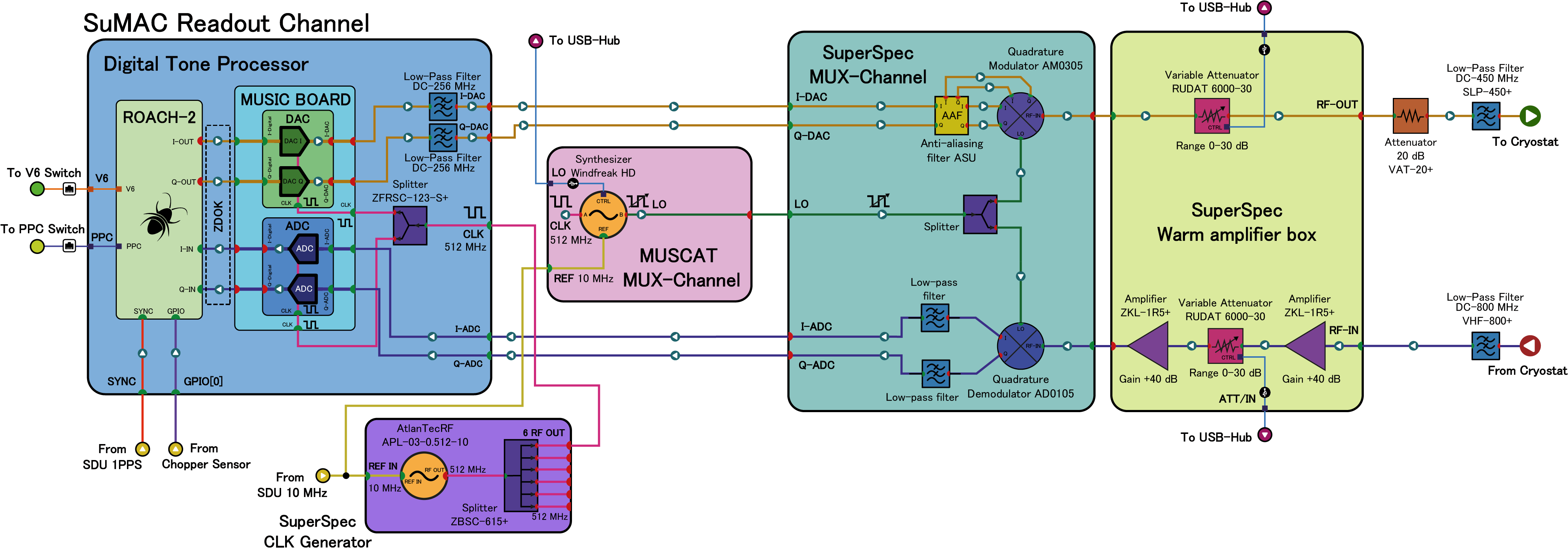}
\end{tabular}
\end{center}
\caption[example] 
{ \label{fig:fig4} Electrical diagram of a SuMAC readout channel. We retain the MUSCAT infrastructure for the processing, generation, and analogue-to-digital conversion of the tone comb (ROACH-2 + MUSIC board). The clock signal is provided by the SuperSpec generator. For up- and down-conversion and the attenuation/amplification stage to and from the cryostat, we primarily use SuperSpec hardware, except for the local oscillator signal, which is obtained from the MUSCAT MUX channel.}
\end{figure} 
    
    \item For SuMAC, we designed and manufactured a chopping mechanism consisting of a two-blade aluminium honeycomb wheel positioned in front of the MUSCAT cryostat window (see Figure \ref{fig:fig5}e). By adjusting the input voltage to the DC motor coupled to the wheel's main shaft, we rotate the blades at 5 Hz, resulting in an effective modulation frequency of 10 Hz. An optical sensor placed at the edge of the wheel generates a pulse each time a blade passes the window. This signal is continuously and simultaneously recorded by each of the ROACH-2 digital inputs.

\end{itemize}

Figure \ref{fig:fig5} depicts various stages of the SuMAC installation in the LMT receiver room. We began by installing the SuperSpec array at the MUSCAT focal plane, along with the low-noise amplifiers, and replacing/repairing coaxial cables. Subsequently, we closed the cryostat, positioned it, and initiated cooling, while simultaneously assembling the SuMAC readout system. Finally, we installed the chopper wheel and its support structure and aligned the warm mirrors with the instrument.

Operating SuperSpec in MUSCAT presents both advantages and disadvantages regarding performance compared to SuperSpec's native cryostat. First, because MUSCAT is equipped with a dilution refrigerator, SuperSpec's operating temperature under typical conditions is $\sim$95 mK — well below the 250 mK of the original SuperSpec deployment cryostat. This significant temperature reduction substantially lowers detector generation-recombination noise, thereby enhancing the instrument's sensitivity. Furthermore, as quality factors increase, the likelihood of resonator collisions decreases, improving the array yield. However, this temperature reduction can also exacerbate Two-Level System (TLS) noise, thereby worsening 1/f noise~\cite{Kumar2008}.

Moreover, the MUSCAT optics are not optimised for SuperSpec at the LMT, thereby reducing observing efficiency and the sensitivity of the SuMAC-LMT system. Furthermore, although we installed a chopping wheel for the SuMAC system, it is not as efficient as the position chopping system designed for the original SuperSpec optics. Consequently, SuMAC requires moving the telescope to observe a point source with the three focal-plane pixels, thereby impacting observation efficiency\footnote{Originally, a system of mirrors was designed for the SuperSpec cryostat to chop the telescope beam across three spatial pixels on the focal plane without moving the telescope itself. 
Observing efficiency is affected only if a ``pointing-switching'' strategy is used and the source extent is smaller than the on-sky pixel spacing.}.

\begin{figure} [ht]
\begin{center}
\begin{tabular}{c}
\includegraphics[height=10.1cm]{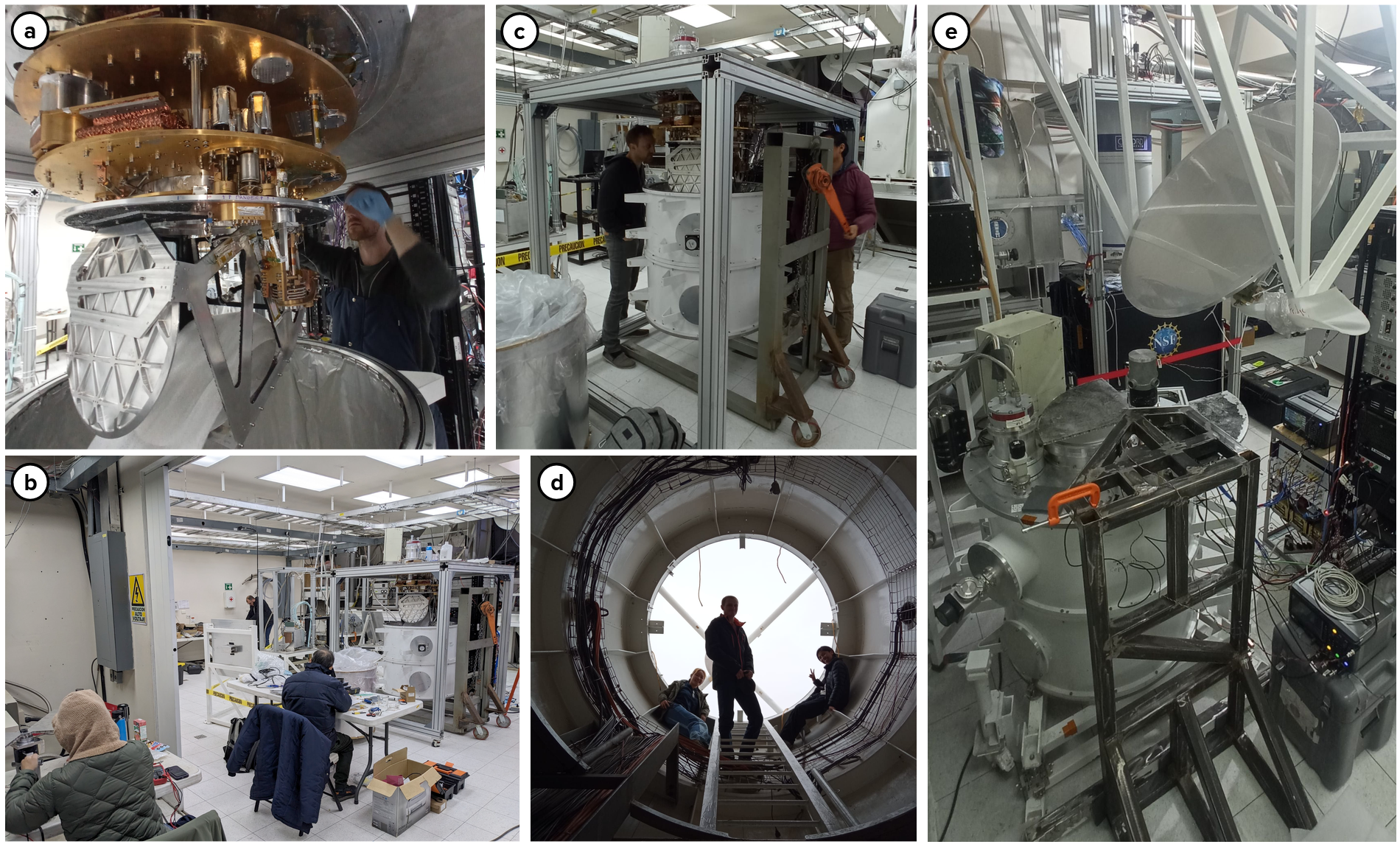}
\end{tabular}
\end{center}
\caption[example] 
{ \label{fig:fig5} SuMAC installation at the LMT during the summer of 2025. a) Installation of the SuperSpec array at the MUSCAT focal plane. b) Testing of the chopper wheel rotation, along with LNA wiring and installation. c) Sealing of SuMAC for on-site placement and start cooling down. d) View of the LMT apex from the receiver cabin. e) Final on-site installation of SuMAC for observations, with the chopping wheel secured to the side by a metal structure.}
\end{figure} 

\begin{figure} [ht]
   \begin{center}
   \begin{tabular}{c}
   \includegraphics[height=10cm]{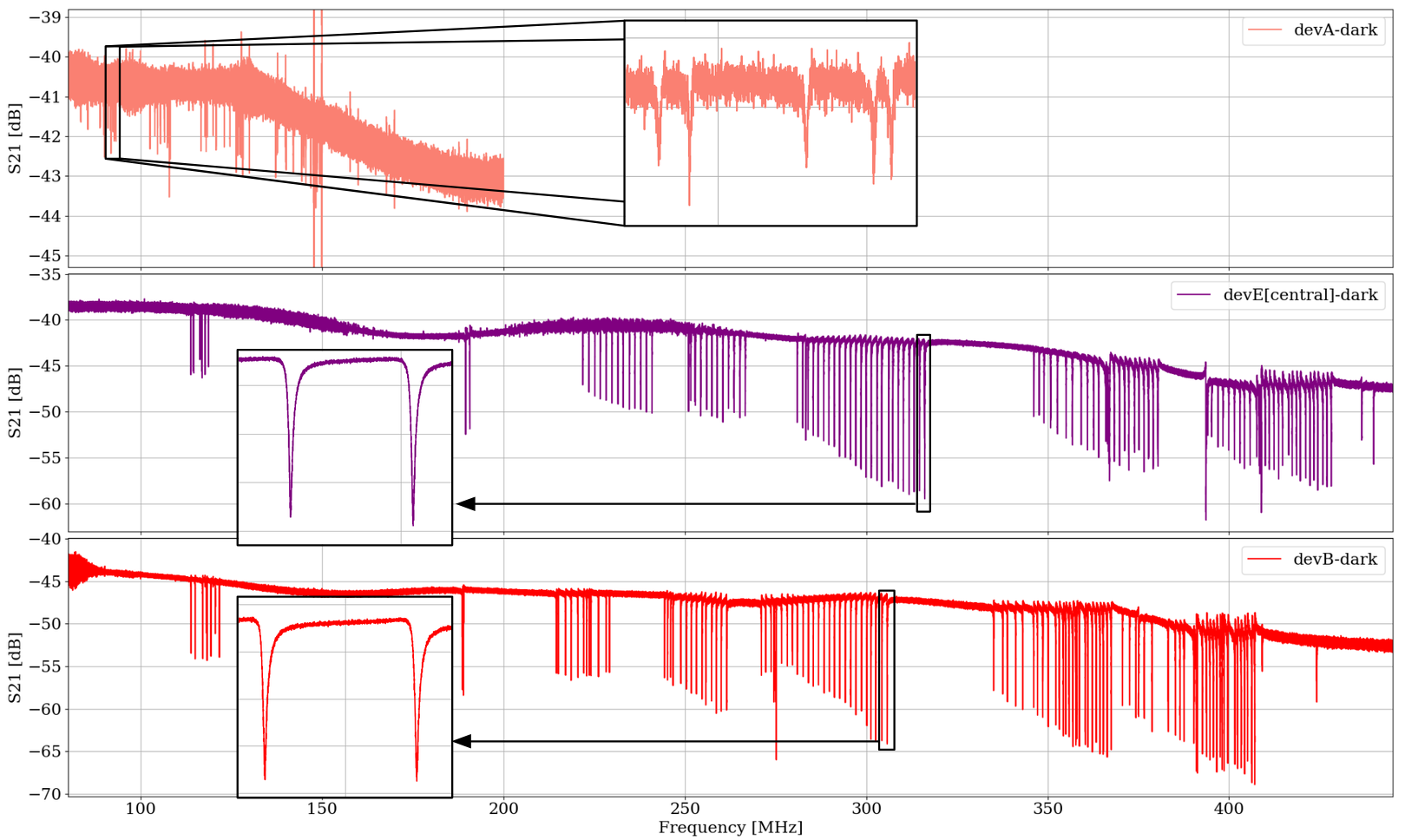}
   \end{tabular}
   \end{center}
   \caption[example] 
   { \label{fig:fig6} Frequency sweeps of the $S_{21}$ parameter for each of the three SuperSpec pixels in MUSCAT at 90 mK with the window closed (minimal optical load). Top: Side pixel devA; we identify 49 resonators between 84 and 119 MHz, generally shallow and noisy. Middle: Central pixel devE, with 122 resonators between 114 and 440 MHz. Bottom: Side pixel devB, with 122 resonators between 114 and 424 MHz. The drive power is -93 dBm, and a 20 dB gain warm amplifier was added at the output.}
   \end{figure}

Once the base temperature of 90 mK was reached\footnote{Under conditions of minimal optical load with the window closed, the SuperSpec array operates at 90 mK, whereas under the optical load from the sky with the window open, the temperature rises slightly to 95 mK.}, we performed a frequency sweep of the $S_{21}$ parameter for each of the SuperSpec pixels using a Vector Network Analyzer (VNA), obtaining the results of Figure~\ref{fig:fig6}. For the central pixel, dev E, we identified 122 resonators between 114 and 440 MHz; we also found 122 resonators in dev B (114–424 MHz) and 49 in dev A (84–149 MHz). The drive power for the three pixels was adjusted to approximately -93 dBm.

Pixel dev A showed considerable baseline noise and very shallow resonators. In contrast, dev E and dev B—which share the same number and distribution of detectors—exhibit very similar transmission characteristics with deeper resonators with dip depths ranging from $\sim$6 to $\sim$15 dB. The spectral responses were obtained through laboratory measurements using a Fourier Transform Spectrometer (FTS), as presented in Kane et al. (2026)\cite{Kane2026}.

\section{SuMAC at the LMT}

\subsection{Hot/cold load measurements}

To directly measure SuMAC's performance prior to coupling with the telescope optics, we estimated the responsivity between two known, distinct optical loads by measuring the detector's resonant frequency shift resulting from a known change in load temperature.

\begin{figure} [ht]
   \begin{center}
   \begin{tabular}{c}
   \includegraphics[height=10cm]{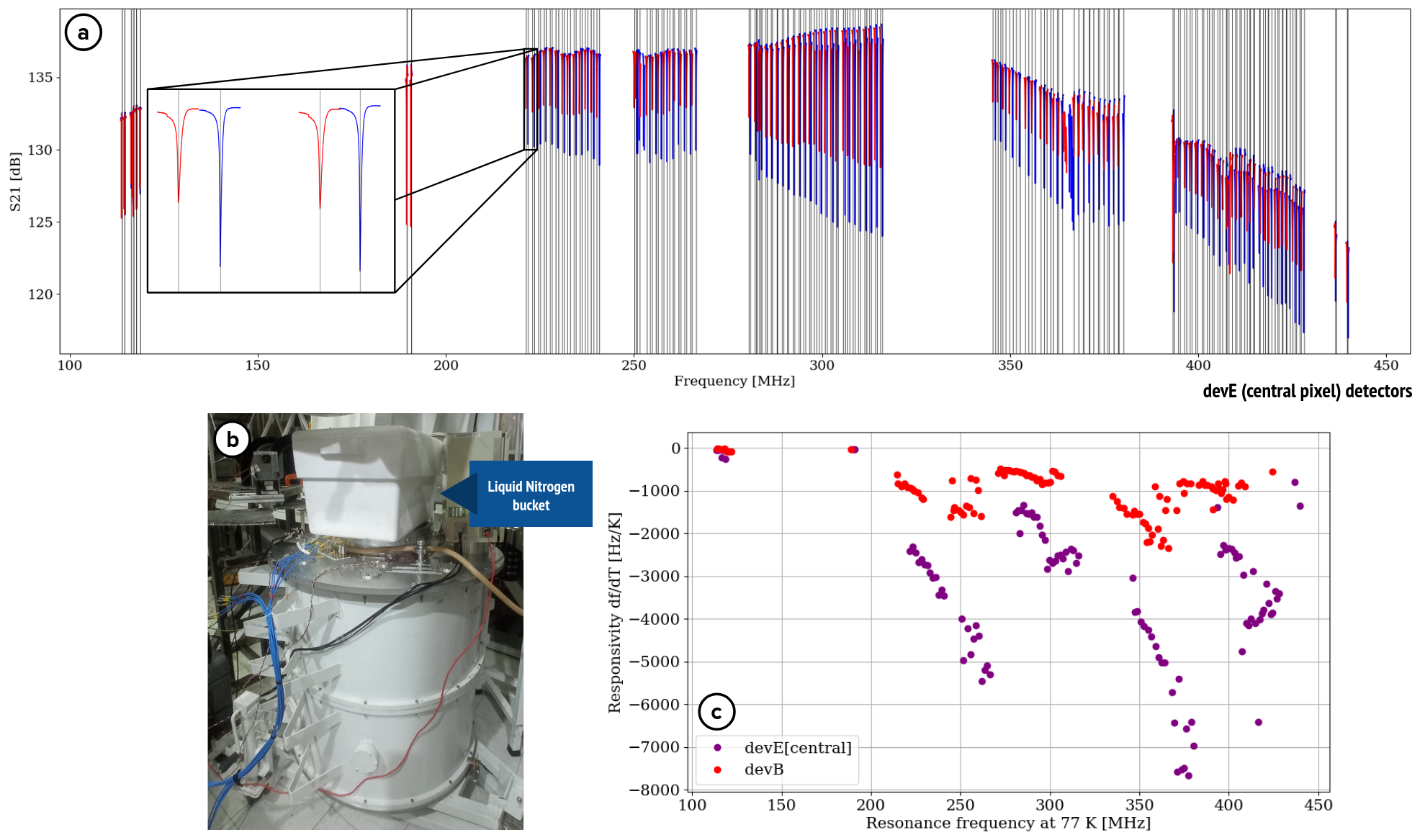}
   \end{tabular}
   \end{center}
   \caption[example] 
   { \label{fig:fig7} SuMAC hot/cold load test. a) Segmented frequency sweep of the central pixel (devE) under two different optical loads: hot at 287 K (red) and cold at 77 K (blue). b) Cold load experiment. A Styrofoam bucket, large enough to cover the window, with walls lined with Eccosorb, is filled with liquid nitrogen. In this setup, frequency sweeps and noise samples are taken for pixels dev E and B. c) Responsivities (df/dT) per detector as a function of their resonance frequencies for dev E (purple) and B (red).}
   \end{figure}

For the cold load, a Styrofoam bucket with walls and bottom lined with Eccosorb was filled with liquid nitrogen (T $\sim77$ K). The container is sized to completely cover the MUSCAT window, as shown in Figure \ref{fig:fig7}b. For the hot load, we covered the window with Eccosorb foam at room temperature—approximately 287 K, according to room temperature sensors. We performed a narrow frequency sweep for each resonator in devices E and B\footnote{The shallow depth from the decreased quality factor caused a significantly reduced readout signal in the device A resonators, making frequency tracking difficult.}. The plot in Figure \ref{fig:fig7}a shows the frequency sweeps for both loads on the device E detectors. The frequency shift and the decrease in the quality factor (depth and width of the resonator response) resulting from the change in optical load are evident. The responsivity for detector $k$, $R_k \mathrm{[Hz/K]}$, is calculated as:

$$
R_k = \frac{f_{0\,k,\rm hot}- f_{0\,k,\rm cold}}{T_{\rm hot} - T_{\rm cold}} = \frac{f_{0\,k,\rm hot}- f_{0\,k,\rm cold}}{287 - 77} = \frac{1}{210} (f_{0\,k,\rm hot}- f_{0\,k,\rm cold})
$$

\noindent where $f_{0\,k,\rm cold}$ and $f_{0\,k,\rm hot}$ are the resonance frequencies for the cold and hot loads, respectively. Figure \ref{fig:fig7}c shows the responsivity of each detector from devices E and B as a function of its resonance frequency under the cold load. In general, we identify three groups: i) dark detectors (not directly coupled to the microstrip feedline) and broadband detectors (weakly coupled over the full frequency range) with expected very low responsivities \cite{Redford_2022}; ii) intermediate; and iii) high, with the latter two categories resulting from a design error that couples optical power to one side of the microstrip line than the other, creating two groups that are interleaved in both optical and readout frequency.

\begin{figure} [ht]
\begin{center}
\begin{tabular}{c}
\includegraphics[height=0.33\textheight]{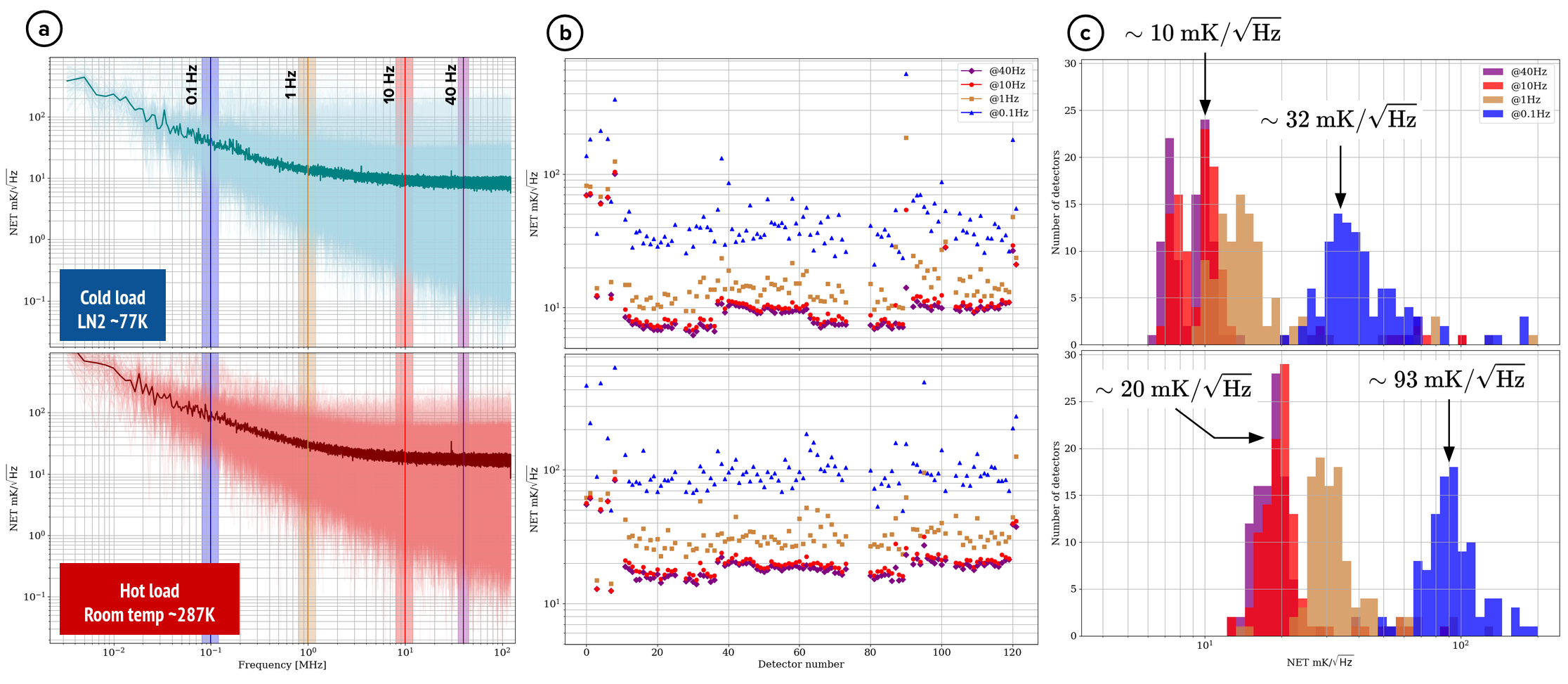}
\end{tabular}
\end{center}
\caption[example] 
{ \label{fig:fig8} a) NET vs sampling frequency for all detectors under cold (top) and hot (bottom) loads. The darker lines show the median of all spectra. The vertical lines and their shaded areas indicate the frequency ranges over which the NET is averaged. b) NET per detector as a function of its detection frequency. Each colour represents the averaged frequency span. c) NET histograms at different frequencies. Under the cold load, the NET ranges from 10 mK/$\mathrm{\sqrt{Hz}}$ in the flat region of the spectrum to 32 mK/$\mathrm{\sqrt{Hz}}$ at low frequencies (0.1 Hz). This condition closely approximates the atmospheric load under good observing conditions (clear skies, opacity below 0.3, and elevations above 30 degrees). In the same range, under a higher load, the NET ranges from 20 to 93 mK/$\mathrm{\sqrt{Hz}}$.}
\end{figure}

For each optical load, we acquired 5 minutes of time-stream data for all detectors. In both cases, the power spectral density (PSD) (see Figure \ref{fig:fig8}a) for a typical detector is flat down to $\sim$3 Hz; below this frequency, 1/f noise dominates and rises rapidly as the frequency decreases. From detector responsivity $R_k$ and the PSD of the noise measurements, $S_{k,ff}$, we can estimate the noise-equivalent temperature per detector, NET$_{k}$, as:

\begin{equation}
    \mathrm{NET}_{k} \mathrm{[K/\sqrt{Hz}]} = \frac{ \sqrt{ S_{k,ff} \mathrm{[Hz^{2}/Hz]} } } { R_k \mathrm{[Hz/K]} }.
\end{equation}

The plots in Figure \ref{fig:fig8}b show the $\mathrm{NET_{k}}$ values as a function of resonance frequency for both optical loads. Each colour corresponds to the NET value for different sampling frequencies: 0.1, 1, 10, and 40 Hz. Figure \ref{fig:fig8}c shows the corresponding histograms. First, we observe that sensitivity worsens—i.e., the NET increases—below 3 Hz due to the strong contribution of 1/f noise; at 0.1 Hz, the NET is three times higher than in the white region under the cold load and up to four times higher under the hot load. The analysis by Redford et al. (2026)\cite{Redford_2022} suggests a strong correlation between detector noise components, making it possible to subtract this noise to significantly flatten the spectrum and improve sensitivity at lower frequencies.

On the other hand, we see that the NET under the hot load is approximately twice as high in the white region of the PSD. This is due to the increase in photon noise and quasiparticle recombination noise. Under good atmospheric conditions at the LMT, we expect an optical load closer to that of the cold load; consequently, the expected sensitivity prior to the telescope optics is close to 10 $\mathrm{mK/\sqrt{Hz}}$ for any sampling frequency above 3 Hz (32 $\mathrm{mK/\sqrt{Hz}}$ at 0.1 Hz).

\subsection{SuMAC commissioning}

From July 18 to August 29, 2025, we commissioned SuMAC on the LMT. A wetter-than-usual Mexican monsoon season limited the observing campaign to 16 available nights in total. Eight of these were morning observations (from 6:00 am local time until sunrise), used primarily to test and improve the Quick-Look Analysis (QLA), communication with the telescope, characterise the beam optimisation procedure, and refine pointing corrections. Of the remaining eight nights, four were partial nights, sharing observation time with other instruments or interrupted by weather conditions, and  four were full nights.

The campaign aimed to calibrate the instrument, characterise its performance, optimise observation parameters and techniques, and verify and evaluate the instrument's capabilities for spectral observations—specifically detecting emission lines, generating multispectral maps, and exploring mapping parameters and techniques for LIM-like observations.

The main observations carried out with SuMAC at the LMT can be classified into four groups:

\begin{enumerate}
    \item Maps. Horizontal raster maps, which can be classified according to their dimensions, as:
    \begin{itemize}
        \item Pointing map. Small 50''$\times$ 50'' maps, 2-3'' steps, and scan rate of 75''/s. To observe point sources using only the central pixel for beam evaluation and pointing error correction.
        \item Beam maps. Large 200'' $\times$ 200'' maps, 2–3'' steps, and a scan rate of 75''/s using point calibration sources. The map dimensions ensure that the source illuminates all three pixels, and the step size is sufficient to resolve the source. We use these maps to characterise the beam shape and perform flux calibration.
        \item Extended source maps. The dimensions, step and scan rate depend on the source size.
    \end{itemize}
    \item Skydips. Noise measurements, frequency and drive power sweeps at various elevations and under specific atmospheric conditions (different opacities). 
    \item Spectrum observations. Observations are made on and off the source to extract atmospheric emission. There are two types: a) single-position switching, shifting the telescope on and off the source at intervals of 10 to 15 seconds, and b) double-position switching, which is similar but involves defining two off-source positions.
    \item LIM-like observations. Azimuth scans at a fixed elevation, with different scan rates. This is an initial proposal to study the feasibility of conducting LIM-type observations with SuMAC on the LMT\cite{Savorgnano26}.
\end{enumerate}

We repeatedly verify beam quality by observing bright point sources. In the event of beam broadening or distortion, we run standard optimisation routines for focus and primary reflector surface (astigmatism) \cite{TapiaThesis}. The data reduction procedure for the observations presented in this work is described in Lapuente et al. (2026)\cite{Lapuente26}.

\section{Results}

\subsection{Beam shape}

We characterised the beam shape based on observations of bright point sources—specifically the planet Neptune and the quasars 3C84 and 3C454.3. We selected channels meeting the following criteria: a signal-to-noise ratio greater than 5; no frequency overlap with other resonators (a minimum separation of 5 line widths); not overdriven; and a spectral response known with high certainty. This resulted in the selection of 77 out of the 122 channels identified in the frequency sweep shown in Figure \ref{fig:fig6}, covering an effective range of 198 to 296 GHz.

\begin{figure} [ht]
   \begin{center}
   \begin{tabular}{c}
   \includegraphics[height=7cm]{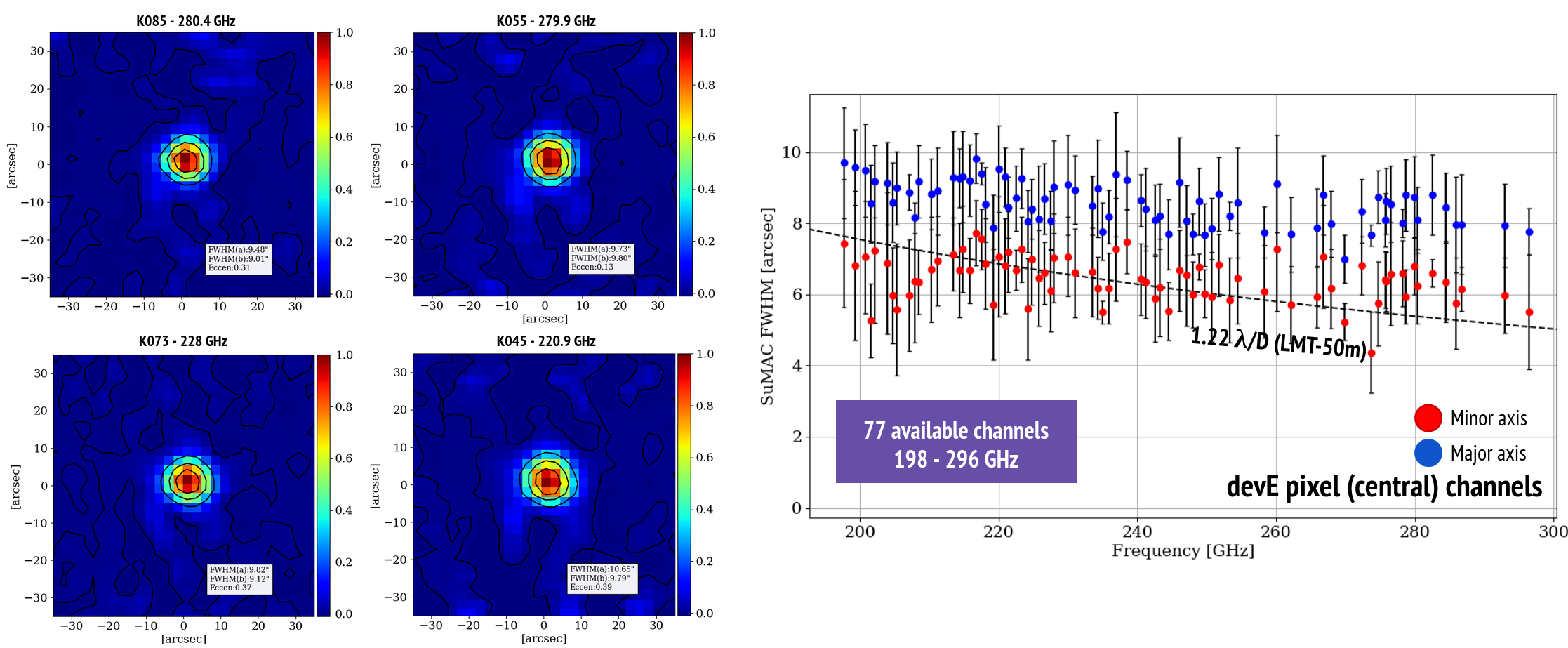}
   \end{tabular}
   \end{center}
   \caption[example] 
   { \label{fig:fig9} SuMAC beam shape results. Left: Maps of multiple point sources coadded for some of the central pixel channels. The beams are round with a slight elongation. Maps processed with a 1.88" Gaussian filter. Right: Major (blue diamonds) and minor (red diamonds) FWHMs averaged over a dozen point-like observations based on their detection frequency. In general, all beams present a slight elongation and broadening with respect to the diffraction limit (assuming a 50~m antenna). The beams follow the $\lambda/D$ trend.}
   \end{figure}

First, by fitting a 2D Gaussian function to each of the observations (10 in total) and averaging the results for the major and minor axis widths across the channels of the central pixel (devE), we obtained the plot shown on the right in Figure \ref{fig:fig9}---the beams generally have FWHMs between 5--10''. We observe that the beam width tends to narrow as the detection wavelength decreases (frequency increases), following the expected $\lambda/D$ relationship. The persistent difference between the major and minor axes indicates a slight elongation of the beam. Furthermore, we observe that the width is slightly larger than the diffraction limit. Although this limit is not entirely precise—since it assumes uniform illumination of the entire antenna, which is unlikely given that the SuMAC optics are not optimised for the LMT—it establishes a lower bound; our measurements are close to this value, suggesting reasonably good optical coupling.

Additionally, the maps on the left in Figure \ref{fig:fig9} show the normalised co-addition of all these observations for the selected channels (77) in the devE pixel. In most cases, the beam width follows the $\lambda/D$ relationship and exhibits a rounded shape. However, a pair of sidelobes is visible to the southeast and southwest of the main beam, with intensities representing approximately 10\% of the beam's amplitude.

\subsection{On-sky sensitivity}

Based on the flux calibration described in Lapuente et al. (2026)\cite{Lapuente26}, using observations of Neptune, we estimate the instrument's on-sky sensitivity. We calculate the NET from the timestreams and subsequently from the maps.

\begin{figure} [ht]
   \begin{center}
   \begin{tabular}{c}
   \includegraphics[height=13cm]{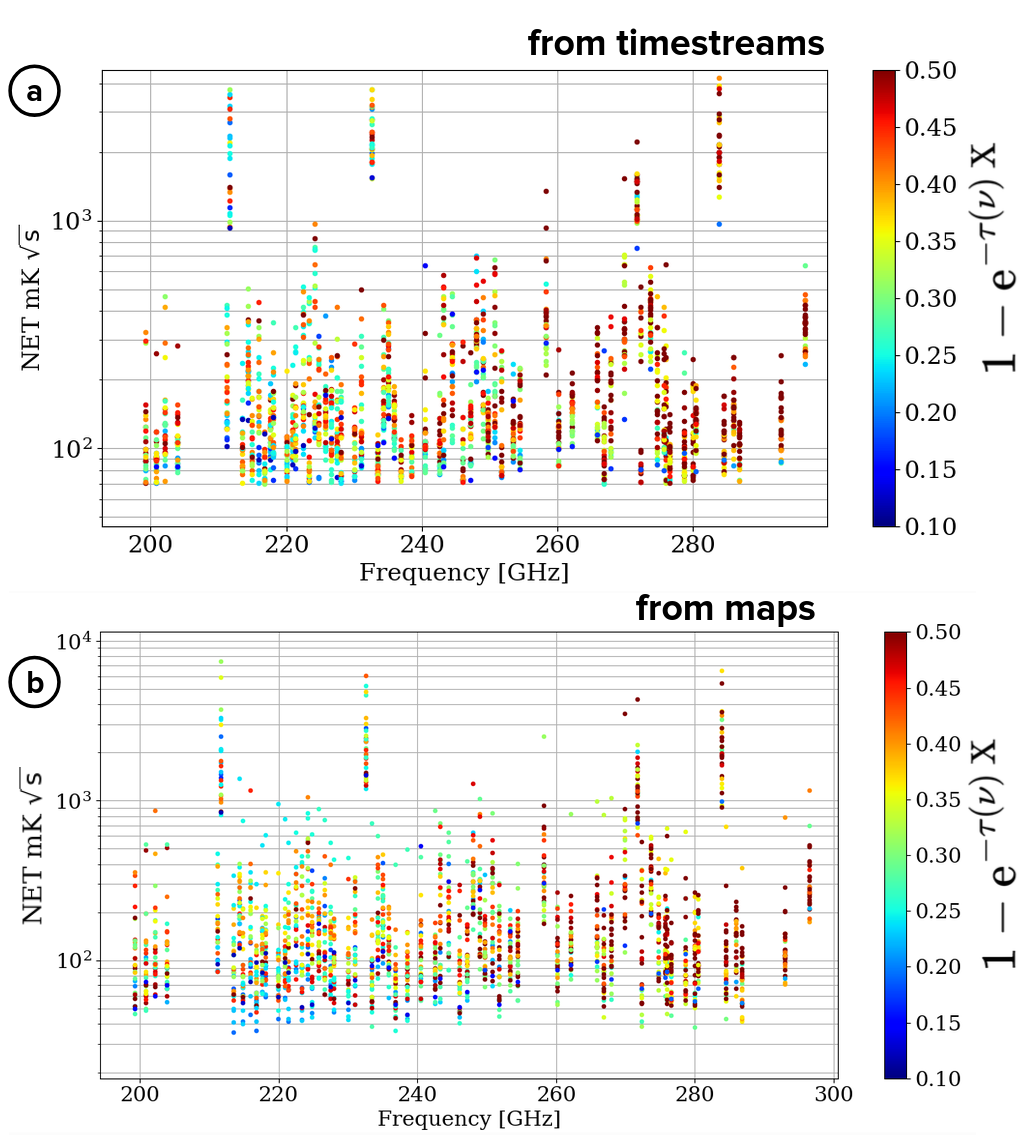}
   \end{tabular}
   \end{center}
   \caption[example] 
   { \label{fig:fig10} SuMAC on-sky sensitivity. a) NET per detector as a function of detection frequency (mm-wavelength), calculated from the timestreams. Each colour represents the level of atmospheric extinction, which is related to the optical load. A slight increase in NET is observed at higher optical loads due to increased wave noise. The median NET value per detector ranges from $\sim$80 to 318 mK $\sqrt{\mathrm{s}}$ with a median value of 106 mK $\sqrt{\mathrm{s}}$. b) NET per detector derived from the maps. Overall, the results from both methods are consistent, with the median NET values using this method ranging from $\sim$70 to 360 mK $\sqrt{\mathrm{s}}$ and a median value of 115 mK $\sqrt{\mathrm{s}}$.}
   \end{figure}

From the timestreams, we can estimate the NET using the definition in Equation 1, expressing the responsivity in terms of the per-detector gains, $G_k$ (estimated in Lapuente et al. (2026)\cite{Lapuente26}), as $R_k = G_k^{-1}$ \footnote{Since the gains are derived from point-source observations, a beam dilution correction must be applied.}. From the resulting NET-versus-sampling-frequency curve, we average the values within a flat region of the PSD, from 5 to 10 Hz. Figure \ref{fig:fig10}a shows the per-detector results obtained from various point-source observations conducted under different atmospheric conditions and at different elevations. Each colour represents the degree of atmospheric extinction, defined as $1 - e^{-\tau_{\nu} X}$, where $\tau_{\nu}$ is the atmospheric opacity at a given frequency and $X$ is the airmass. This quantity reflects the magnitude of the optical loading on SuMAC. The plot reveals a slight increase in NET as observations are made under higher optical loading, as expected, due to the increase in wave noise. Considering the full range of atmospheric conditions that prevailed during our observation campaign, the typical NET per detector (median) ranges from $\sim$80 to 318~mK~$\mathrm{\sqrt{s}}$, with some outliers well above this range. The typical value across the detectors (median of medians) is 106 mK $\sqrt{\mathrm{s}}$.

To estimate the NET from the maps, it is necessary to reduce the raw data and generate the maps as described in Lapuente et al. (2026)\cite{Lapuente26}. For each signal map per detector, we also generate the corresponding coverage map. We calculate the NET per detector as $\mathrm{NET}_{k} = \sigma_k~\sqrt{t_{k,\,\rm int}}$. $\sigma_k$ is the standard deviation of the signal map in a region away from the source, which we select as a ring centred on the source with inner and outer diameters of 40" and 160", respectively. Here, $t_{k,int}$ is the average integration time on the coverage map within the region selected for noise calculation. Figure \ref{fig:fig10}b shows the NET values per detector extracted from the maps for the same observations as in Figure \ref{fig:fig10}a. The results are similar to those obtained from the timestreams; the median values per detector range from $\sim$70 to 360 mK $\sqrt{\mathrm{s}}$, with a typical value across the detectors of 115 mK $\sqrt{\mathrm{s}}$. A degradation in sensitivity with the optical load due to wave noise is also observed.

Comparing the NET at the cryostat aperture under a 77 K load (NET = 10 mK/$\mathrm{\sqrt{Hz}}$ = 7.1 mK $\mathrm{\sqrt{s}}$) with the average on-sky NET value of a typical observation, of 106 mK$\mathrm{\sqrt{s}}$ (from timestreams), we note a substantial reduction in sensitivity due to several factors: the sky loading being much higher than the cold load, non-optimized coupling to the optics, losses from the four mirrors internal to the cabin, antenna shape (time-variable) and surface roughness, atmospheric conditions and uncertainty in the opacity\cite{Lapuente26}.

\subsubsection{SuMAC spectroscopic capabilities}

\begin{figure} [ht]
   \begin{center}
   \begin{tabular}{c}
   \includegraphics[width=1\textwidth]{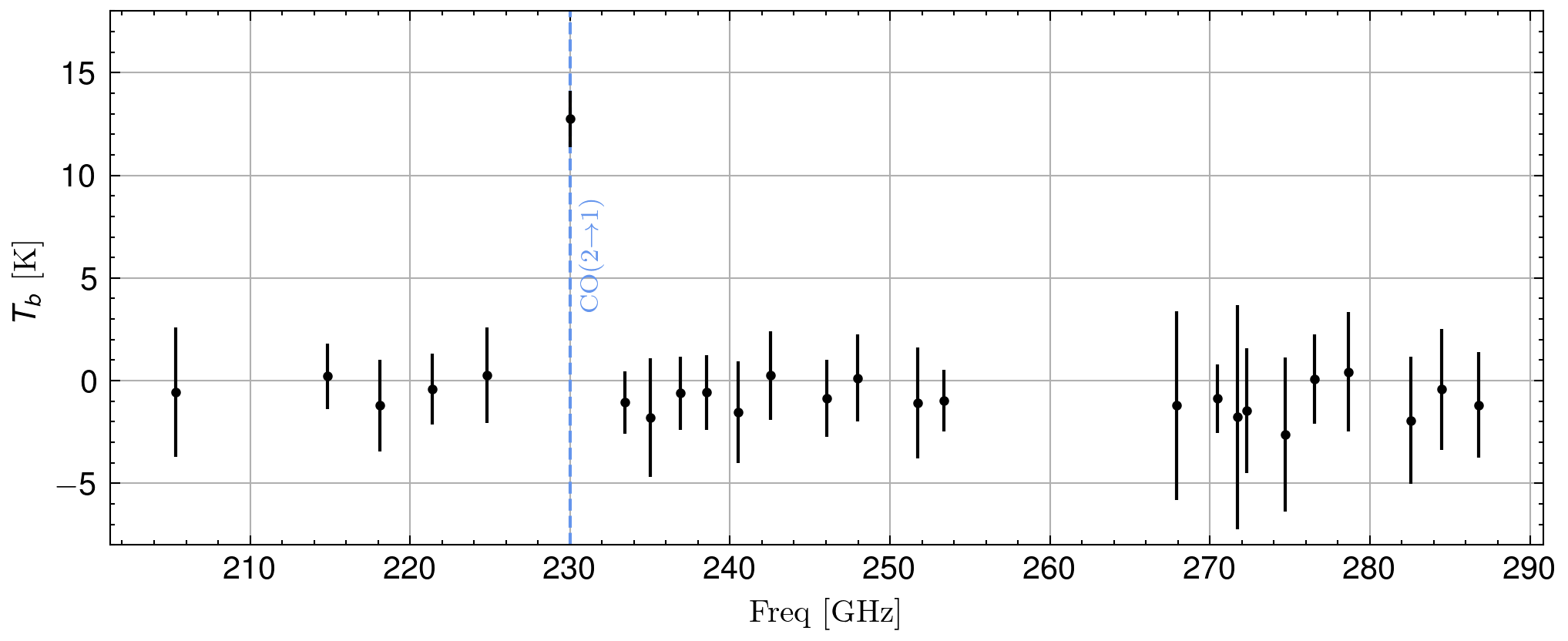}
   \end{tabular}
   \end{center}
   \caption[example] 
   { \label{fig:figSpectra} Example preliminary spectrum of starburst galaxy NGC 253 observed by the central pixel.}
   \end{figure}

To verify SuMAC's spectroscopic capabilities, we observed the centre of the spiral galaxy NGC253 using the central pixel and the `pointing-switching' technique. Figure \ref{fig:figSpectra} shows the reduced SuMAC spectrum, highlighting the detection of the CO($J=2\rightarrow1$) emission line at 230.5 GHz—previously detected by instruments such as IRAM and JCMT—which reveals active star formation and outflow structures in its core \cite{Harrison1999}. The observation procedure and data reduction are described in greater detail in Lapuente et al. (2026)\cite{Lapuente26}.

\begin{figure} [ht]
   \begin{center}
   \begin{tabular}{c}
   \includegraphics[height=13cm]{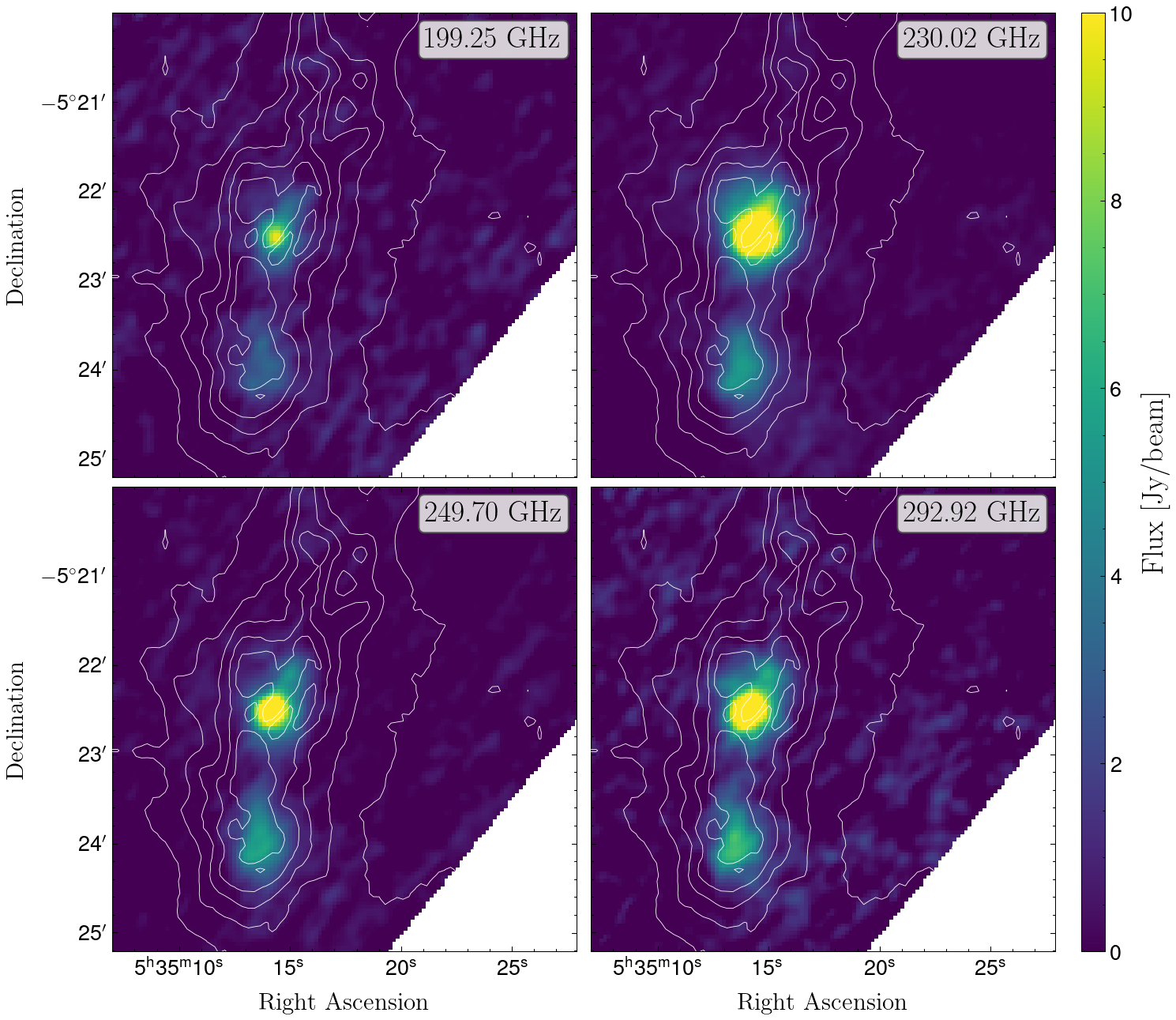}
   \end{tabular}
   \end{center}
   \caption[example] 
   { \label{fig:figOrion} The Orion KL nebula at four frequencies across our observing bandwidth of the central pixel (devE). The nebula's hot core can be seen, as well as its fainter southern lobe. White contours show the JCMT SCUBA-2 850~$\mu m$ continuum emission at significance levels of $50$, $100$, $150$, $200$, and $250\sigma$, where $\sigma$ is the local RMS noise derived from the map's per-pixel variance\cite{WardThompson07,Kirk18}.}
   \end{figure}

To demonstrate the instrument's capability for multispectral mapping of objects or regions, we observed the Orion Molecular Cloud 1 (OMC-1) within the Orion Nebula, a region of massive star formation in the solar neighbourhood. We performed 12 raster-scan maps over a 6'' $\times$ 6'' area centred on the Kleinmann-Low (KL) Nebula. Figure \ref{fig:figOrion} displays the resulting maps from some selected channels from the central pixel (devE). The frequency dependence of the intensity and shape of the nebula's brightest regions is notable; the region appears particularly bright in the 230 GHz channel, where significant CO($J=2\rightarrow1$) emission has been detected\cite{Zapata_2011}. For comparison, contours from the SCUBA-2 map (JCMT) at 850 µm are also shown. Morphological agreement is observed between the brightest structures detected by both cameras. Lapuente et al. (2026) \cite{Lapuente26} detail the observation parameters and the reduction procedure.

\section{Conclusions}

We successfully installed, integrated, and operated a three-pixel SuperSpec array at the focal plane of MUSCAT at the LMT. Compared to the originally-planned deployment in a different cryostat, the intrinsic detector noise was lower due to MUSCAT's dilution refrigerator operating at 95 mK vs 250 mK\cite{Karkare_2020}. However, the optical system was not optimised for the SuperSpec pixels and lacked a beam-chopping system, which counteracted the sensitivity gains from lower-temperature operation. Two 110-channel devices performed well, with deep resonances and high signal-to-noise measurements. 
Detector responsivity was measured directly at the SuMAC window by calculating the frequency shift resulting from the change between a cold (77 K) and a hot (300 K) optical load. Using noise stares we estimated the NET for both loads, obtaining typical values of 10 and 20 mK$/\sqrt{\text{Hz}}$ for the cold and warm loads, respectively. Typical sky loadings at the LMT are in between the two (but closer to the cold load). 

We conducted the SuMAC observation campaign at the LMT over a period of just over a month. During the few nights when conditions permitted, we carried out various observations using different techniques to characterise and calibrate the instrument. We estimated the on-sky NET per detector for the central pixel using both timestreams and maps, finding strong consistency between the two. Although sensitivity depends on atmospheric observing conditions, for a typical summer night at the LMT, we estimated a mean value of 106 mK $\sqrt{\mathrm{s}}$ (115 mK$\sqrt{s}$ from maps). This reduction in sensitivity compared to that measured at the cryostat aperture is attributable to factors such as suboptimal optical coupling, losses at the four warm mirrors in the cabin, antenna surface roughness, and atmospheric uncertainties\cite{Lapuente26}, and is roughly consistent with values measured from other LMT instruments.

Our preliminary observations demonstrated SuMAC's spectral mapping capabilities, detecting a spectral line in the galaxy NGC253 and making multispectral maps of OMC-1 with the central pixel (devE). We also produced LIM-like maps—azimuth scans at a fixed elevation centered on galaxy clusters—to study noise properties while various observational parameters were varied in a scanning mode not typically used at the LMT\cite{Savorgnano26}.

As the first deployment of on-chip spectrometers on a large millimeter-wave telescope, SuMAC is a pathfinder for future, higher-density focal planes which will allow large-scale surveys of high-redshift, dusty galaxies. More detailed analysis of our data, including the second pixel, will feed into the design of next-generation spectroscopic instruments.


\acknowledgments 

This work is based on observations obtained with the Large Millimeter Telescope Alfonso Serrano, a binational project of the Instituto Nacional de Astrofísica, Óptica y Electrónica (INAOE) and the University of Massachusetts Amherst. We thank the LMT staff for their support during the installation and observing campaign, as well as Dr.\ Hien Nguyen and Leslie Barrios for their invaluable support in critical stages of the project. This project was possible due to partial support from SECIHTI research grant CBF-2026/4130.

\bibliography{report} 
\bibliographystyle{spiebib} 

\end{document}